\documentclass[onecolumn,10pt,showpacs,preprint]{revtex4-1}
\usepackage{graphicx,anysize,amsmath,amssymb,mathrsfs}
\usepackage{bbm,bm,epstopdf}
\usepackage{multirow}
\usepackage{threeparttable}
\usepackage{float}
\usepackage{epic,eepic}
\usepackage{makeidx}
\usepackage{epsfig}
\usepackage{color}
\usepackage{subfigure}
\usepackage{supertabular}
\usepackage{hyperref,varioref}
\hypersetup{colorlinks,citecolor=blue,linkcolor=red,urlcolor=blue}

\begin{document}
\bibliographystyle{prsty}
\title{Organelle morphogenesis by active remodeling}

\author{N. Ramakrishnan$^{1,2}$}
 \email{ramn@seas.upenn.edu}
 \author{John H. Ipsen$^3$}
\email{ipsen@memphys.sdu.dk}
 \author{Madan Rao$^{4,5}$}
 \email{madan@ncbs.res.in}
\author{P. B. Sunil Kumar$^{1}$}
\email{sunil@physics.iitm.ac.in}
\affiliation{$^1$ Department of Physics, Indian Institute of Technology Madras, Chennai, India, 600036}
\affiliation{$^2$ Department of Bioengineering,  University of Pennsylvania, Philadelphia, PA-19104}
\affiliation{$^3$MEMPHYS- Center for Biomembrane Physics, Department of Physics and Chemistry, University of Southern Denmark, Campusvej 55, DK-5230 Odense M, Denmark } 
\affiliation{$^4$Raman Research Institute, Bengaluru 560080, India}\affiliation{$^5$National Centre for Biological Sciences (TIFR), Bengaluru 560065, India}

\date{\today}
\begin{abstract}
 Intracellular organelles are subject to a steady flux of lipids and proteins through active, energy consuming transport processes. Active fission and fusion are promoted by GTPases, e.g., Arf-Coatamer and the Rab-Snare complexes, which both sense and generate local membrane curvature. Here we investigate through Dynamical Triangulation Monte Carlo simulations, the role that these active processes play in determining the morphology and  compositional segregation in closed membranes. Our results suggest that the ramified morphologies of organelles observed in-vivo are a consequence of driven nonequilibrium processes rather than equilibrium forces. 
\end{abstract}
\maketitle

A characteristic feature of eukaryotic cells is the variety of membrane bound organelles, 
distinguished by their unique morphology and chemical composition. These internal organelles emerge in the backdrop of  a steady flux of material (lipids/proteins) carried by membrane bound vesicles which fuse into and fission off from them. A central issue in cell biology is to explain the morphology and chemical composition of organelles as a consequence of the molecular processes and physical forces involved in this transport \cite{Rafelski:2008iy,Marshall:2011kw,MartinezMenarguez:2013ek}. While there is a rather detailed knowledge of the molecular processes involved in membrane remodeling, our understanding of the underlying physical principles is still quite rudimentary 
\cite{Shibata:2009je}.
 
 To arrive at these underlying principles, one approach is to construct theoretical models describing both the transport process and membrane morphology and to to have them contend with high resolution experiments on live cells. This might appear as a daunting task, if only because any `realistic' model would contain a large number of undetermined parameters. On the other hand, as has been demonstrated in many theoretical studies \cite{Rao:2001jl,Girard:2004jl,Klann:2012ew,Ispolatov:2013bc}, a useful strategy is  to construct coarse-grained theoretical models, described by a few generic, agreed upon rules with minimal  molecular detail.
 
An incontrovertible aspect of cellular organelles is that they are dynamic membranous structures, subject to and driven by a continuous  flux of membrane bound material \cite{Alberts:1994}. Several studies have shown that the time scales of material flux via vesicle fission and fusion onto a compartment \cite{Wieland:1987tw} are at least comparable to membrane relaxation times, which for a micron sized compartment is of the order of tens of seconds. Thus the large scale morphology of the membrane bound compartments must be influenced by active out-of-equilibrium processes of fission and fusion of material.

\begin{figure}[!h] 
\centering
\includegraphics[width=3.0in]{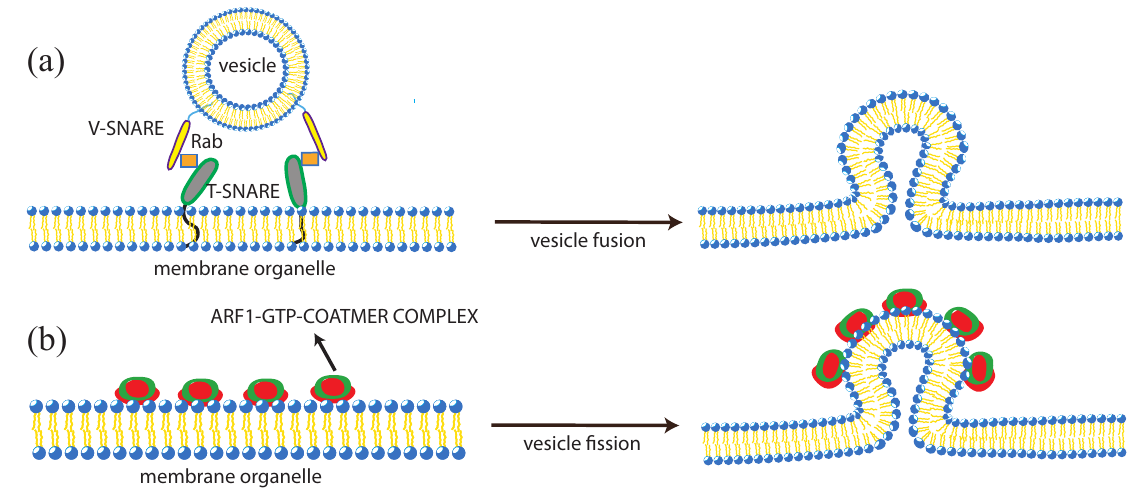}
\caption{\label{fig:prot-bound-unbound} {\bf Active mechanisms of local curvature induction.} (a) The Rab-Snare complex binds at the transport vesicle-organelle interface and induces 
membrane deformation and fusion. (b)  The Arf-Coat complex binds to the membrane organelle and induces 
membrane deformation and fission of a transport vesicle. In addition to these active processes, we describe the active processes and shape changes following the switching of membrane bound pumps
from their active to inactive forms (Supplementary).
}
\end{figure}

Furthermore, organelles are subject to the action of curvature sensing and curvature generating proteins which modulate local  membrane shape -- such proteins now include a variety of bar-domain proteins \cite{Frost:2009gu}, coat-proteins \cite{DSouzaSchorey:2006gc}  and GTPases \cite{Marks:2001em,Baschieri:2012ia,Harris:2011db}  and are  found on most membrane bound organelles and the plasma membrane. In particular, proteins complexes such as Rabs-Snare and the Arf-Coatamer that promote fusion and fission, respectively \cite{Chavrier:1999ir}, shuttle between membrane bound/unbound states - the mechanochemistry of these bound complexes suggest that they respond to and drive changes in the local curvature of the membrane upon energy consumption \cite{Turner:2005dt}, see Fig. ~\ref{fig:prot-bound-unbound} for a schematic.

In general of course,  vesicle fission-fusion involve changes in both local membrane curvature and membrane area  \cite{Rao:2001jl,Staykova:2011cb}.  In this manuscript, for simplicity,  we consider a  {\it perfectly balanced membrane} where the rates of  fission and fusion are the same - in this limit we ignore fluctuations of {\it lipid number}.
We study of the morphological changes of a fluid vesicle induced by active curvature fluctuations arising from fission-fusion, using a Dynamic Triangulation Monte Carlo (DTMC) simulation. 
This  ignores the effects of hydrodynamics and treats the membrane within a Rouse description.  We display the steady state membrane shapes and a phase diagram as a function of activity rate and the extent of curvature generation per active event  and show how membrane activity manifests as an effective pressure or tension. We conclude with a discussion on the significance of such activity driven membrane remodeling  in describing the shape of intracellular membrane organelles in-vivo.

\section{Model}
Since we are interested in the dynamics of shape changes over large spatial and temporal scales (size of organelle, $10\,\mu$m $\gg$ size of transport vesicle, $100$\,nm; membrane relaxation time scales, $1-10$\,s
$\gg$ inverse rate of material flux), it is appropriate to use a coarse-grained description of the membrane dynamics. The membrane dynamics is then governed by elastic membrane stresses, viscous dissipative stresses
and activity.

\noindent{\it Elasticity of membranes : }
The membrane organelle, treated as a closed elastic sheet,  is assumed {\it tensionless} in the absence of any activity - the elastic stresses can be described by the Canham-Helfrich energy functional,
\begin{equation}
\label{eq:Helf}
\mathscr{H}_{el}=\oint d{\bf s} \left(\frac{\kappa}{2}\left[2H({\bf s})-H_0({\bf s})\right]^{2}+\kappa_{G}K \right)-\Delta p \int dV, 
\end{equation}

where $d{\bf s}$ is the membrane area element and ${\bf s}=(s_1,s_2)$  are the membrane in-plane coordinates.  The local curvature is measured in terms of the local mean curvature $H$ and Gaussian curvature $K$ \cite{doCarmo:1976}. The associated bending moduli $\kappa, \kappa_{G}$ are material parameters and are taken to be uniform. The spontaneous curvature term $H_{0}$ is a measure of the preferred local mean curvature of the membrane and is of relevance only at the sites of activity. We will declare the form of this term later.  In addition, there is an osmotic pressure difference $\Delta p$ which sets the scale of the mean enclosed volume at equilibrium.  In this manuscript we drop the dependence on $K$, since we consider surfaces with fixed spherical topology.
\begin{figure}[!h]
\centering
\includegraphics[width=3.0in]{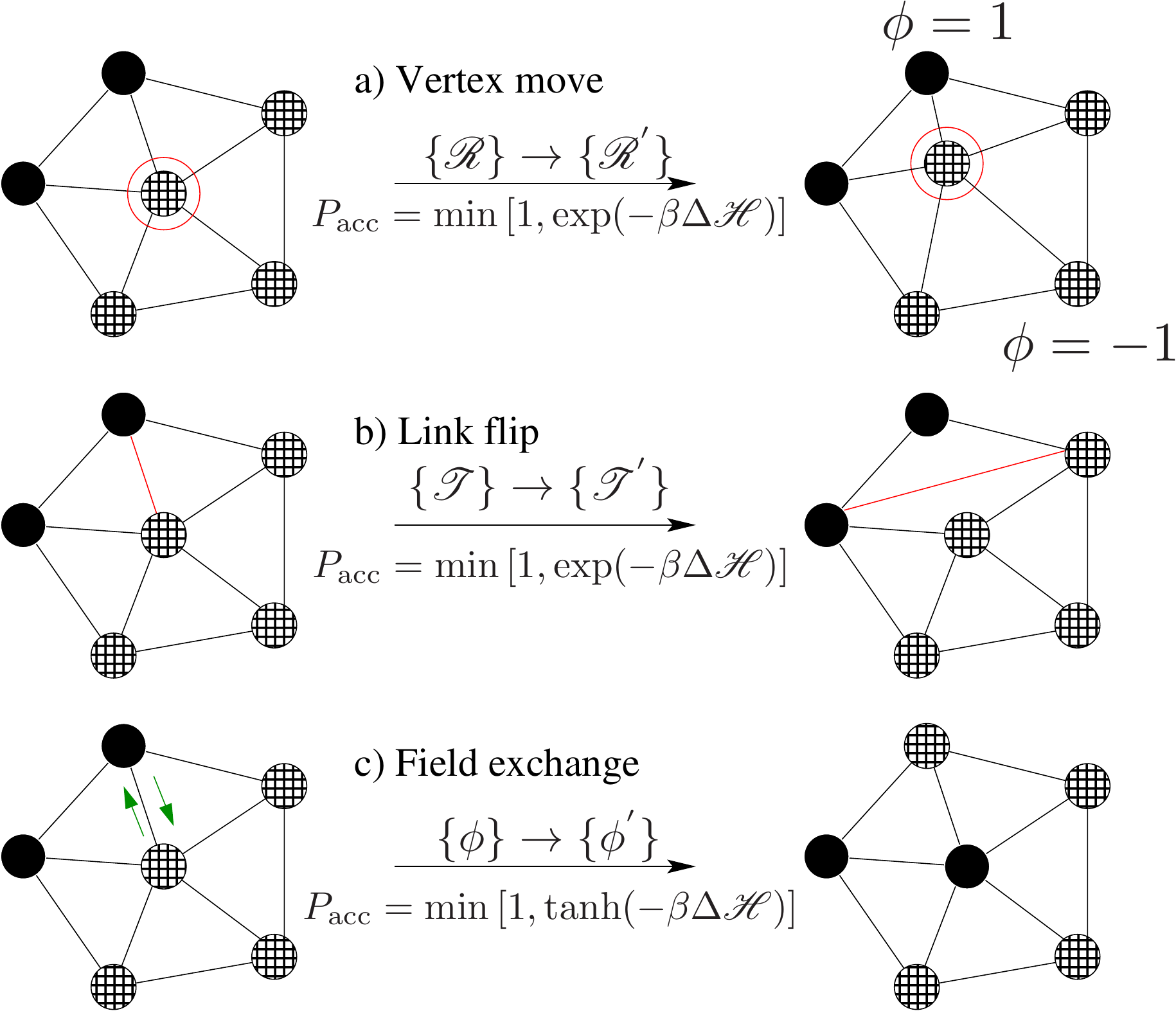}
\caption{\label{fig:mcsmoves} {\bf DTMC of two-component fluid membranes.} (a) A chosen vertex is randomly displaced in 3-dimensions 
keeping the connectivity $\{ \mathscr{T}\}$ unchanged.  (b) A link is flipped (red line) to change connectivity.
 (c)  Kawasaki exchange of  $\{\phi\}$ (green arrows)   to enable diffusion of active protein complex on the surface (see text for notation).}
\end{figure}

To be able to describe the ramified, strongly non-axisymmetric shapes displayed by membrane organelles in-vivo, we resort to computer simulations, where the fluid membrane is represented as a triangulated surface, with $N$ vertices, denoted by $\{\mathscr{R}\}$, which are interconnected to form a  triangulation map,  $\{\mathscr{T}\}$, consisting of $T$ triangles and $L$ links ({\it Sec. S2 in S.I.}). We use the discrete form of the energy functional Eq.\ref{eq:Helf},
\begin{equation}
 \label{eq:2field-mem} \mathscr{H}_{\rm el}= \frac{\kappa}{2} \sum_{i=1}^{N} \left ( H_{i} -H_{0i}\right ) ^{2}A_i-\Delta p \,V \, ,
 \end{equation}
where $A_i$ is the area element associated with vertex $i$.
This has proved to be a valuable tool to study  non-axisymmetric, multicomponent and driven vesicles in a variety of contexts \cite{Gompper:1994cs,SunilKumar:2001cr,Noguchi:2005br,Ramakrishnan:2010hk}.

\noindent{\it Active fission-fusion on membranes : } The active events of fission and fusion are promoted by  curvature generating vesicle-protein complexes represented by a scalar field $\phi$ at every site $i$ which take values $+1$ or $-1$, depending on whether this complex is bound or not (Fig. \ref{fig:prot-bound-unbound}). When bound ($\phi_i=1$), the  complex induces a preferred local curvature at $i$, whereas when unbound ($\phi_i=-1$), the membrane reverts to being locally flat at $i$. This is modeled by assigning a local spontaneous curvature at $i$, given by \footnote{In the case of membrane bound pumps which switch from their active to inactive forms, we take $H_{0i}=C_0\phi_i^2(1+\phi_i)/2$ {\it( see Sec. S1 in S.I.)}.}  $H_{0i}=C_0(1+\phi_i)/2$. The binding/unbinding of these complexes could be cooperative, this could be accounted for by an
 {\it Ising}-interaction,
\begin{equation}
\label{eq:2field-self}
\mathscr{H}_{\phi}=-\frac{1}{2} \sum_{i=1}^{N} \sum_{j \in \Omega_i} J_{ij}\, \phi_{i} \phi_{j}.
\end{equation}
where the sum over sites $j \in \Omega_i$, the set of all vertices connected to $i$. While in principle $J_{ij}$ can depend on the local curvature, for simplicity we take it to be homogeneous and equal to $J$.
We consider the case where $J \geq 0$, the equality holding for complexes that are independent and uncorrelated. On the other hand, $J > 0$ corresponds to complex cooperativity - as suggested by the existence of export sites\cite{Brandizzi:2013ik,Spang:2013ev} in the secretory system.  The total energy of the membrane is given by,
\begin{equation}
\label{eq:2field-tot}
\mathscr{H}=\mathscr{H}_{\rm el}+\mathscr{H}_{\phi}.
\end{equation}
The (un)binding and consequent curvature generation are active processes and are incorporated in the Monte Carlo dynamics as additional moves in the semi grand ensemble. We attempt to flip the field $\phi_i \rightarrow -\phi_i$ locally at a rate $\epsilon_{\pm}$, the probability of acceptance is given by~\cite{Im:20030p788},
  \begin{eqnarray}
 \label{eq:probAI} 
&{\cal P}_{\rm acc}^{+-}
= &
  \left(\frac{N_+}{N}\right) {\rm min}\left \{1, \frac{1}{1+\exp(\zeta [N_+-N_{-}-A_0])} \right \} \nonumber \\
\end{eqnarray}
 and,
\begin{eqnarray} 
\label{eq:probIA}
&{\cal P}_{\rm acc}^{-+}
= &  \left(\frac{N_-}{N}\right) {\rm min}\left \{1, \frac{1}{\eta+\exp(-\zeta [N_+-N_{-}-A_0])} \right \} \nonumber \\
\end{eqnarray}

where $\beta \equiv 1/k_B T$ ($k_B$ is the Boltzmann constant and $T$ the temperature) and $\Delta \mathscr{H}$ is the change in $\mathscr{H}$ \eqref{eq:2field-tot} upon flipping. Here, $N_\pm$ are  the instantaneous number of  sites on the membrane with $\phi=\pm 1$,  with  $N=N_{+}+N_{-}$ being fixed; at  steady state this is controlled by the preferred asymmetry parameter ${A}_{0}\equiv N_{+}^{0}-N_{-}^{0}$ and chemical potential $\zeta$. When $C_0 \ne 0$, the  transitions from $-1 \rightarrow +1$ and  $+1 \rightarrow -1$ involve changes in the membrane curvature energy and are active.
However when $C_0 = 0$, we expect these transitions to be microscopic reversible, which is ensured by setting $\eta=\left(2\frac{N_-}{N_+}-1\right)$
in \eqref{eq:probIA}. 

The in-plane dynamics, fluidity and out-of-plane shape dynamics are described by the time evolution of  the triangulated mesh using Metropolis moves summarized in Fig. \ref{fig:mcsmoves} and detailed in the Supplementary. This is augmented by the active Monte Carlo moves described by
 \eqref{eq:probAI} and \eqref{eq:probIA}.
We define a Monte Carlo sweep (MCS) to be $L$ attempts to flip links, $N$ attempts to move vertices, $N_+$ attempts to exchange $\phi_i$ with its neighbours  and  $\epsilon_\pm$  attempts to flip the value of $\phi$  at vertices. Unless stated otherwise, we use  $\epsilon_{+}=\epsilon_{-}=\epsilon$. 

 We fix $N_+^0=0.1N$ ($10\%$ of the particles are declared active) and vary $\epsilon, C_0$ and $J$ to explore the morphology of membranes.  
$\kappa$ and $J$ are in units of $k_B T$, and $C_0$ is in units of  $a_{0}^{-1}$, where $a_{0}$ is the typical size associated with the coarse grained vertices in the membrane, as discussed in SI.
\section{Results and Discussions} 

Motivated by the trafficking dynamics in the secretory pathway \cite{Wieland:1987tw}, we have chosen parameters to be in the strongly nonequilibrium regime where membrane relaxation times are longer than the timescale of activity. Our object is to provide a phase diagram of steady state shapes of the active vesicle;  for comparison we display the catalogue of shapes exhibited by an equilibrium membrane at different values of 
$C_0$ and $J$  ({\it Fig. S4}). We ensure that these correspond to steady state configurations by computing the time series of the elastic energy and cluster size distribution.

\begin{figure}[!h]
\centering
\includegraphics[width=3.0in]{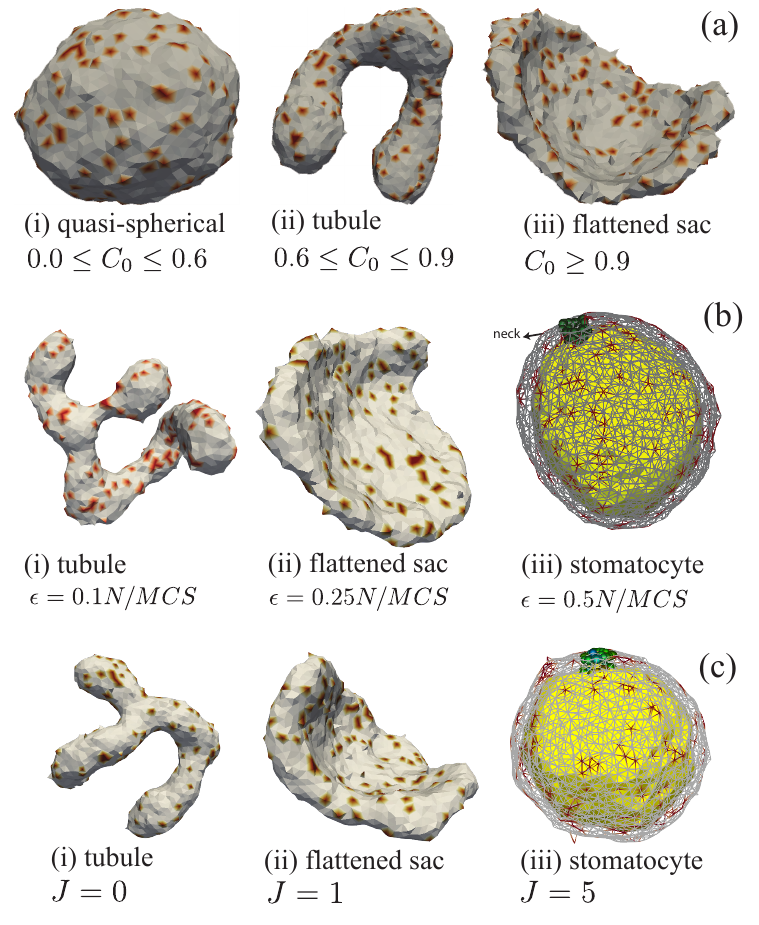}
\caption{ \label{fig:NEq-Rate0.1-J0-conf}{\bf Sequence of shape changes in an active vesicle.}
(a) Steady state shapes of active vesicles at $\epsilon=0.1 N/MCS$ as a function of curvature-activity coupling, $C_0$.
(b) Steady state shapes of active vesicles at  $J=0.0$, and $C_0=0.8$ as a function of activity rate, $\epsilon$. The  side of the stomatocyte that is curved in is colored differently for clarity.
(c) Steady state shapes of active vesicles at $\epsilon=0.1N/MCS$ and $C_0=0.8$ as a function of interaction $J$ between the active species.   In all these configurations, we have fixed $\kappa=20$, $\Delta p=0.0$, and the number of active protein complexes, whose locations are shown by the shaded regions, as $N_{+}^{0}=0.1N$. For comparison, we have also shown the equilibrium shapes of vesicles at different $C_0$ and $J$ in {\it S.I. (Fig.  S5)} 
}
\end{figure} 

\noindent
 \textit{Effect of curvature-activity coupling,\,$C_0$} : 
  Fig. \ref{fig:NEq-Rate0.1-J0-conf}a(i-iii)  shows a sequence of steady state shapes of the active vesicle going from
  quasi-spherical to tubule to flattened sac 
  on increasing $C_0$, at a fixed activity rate $\epsilon=0.1N/MCS$. These shapes are very distinct from the equilibrium vesicle shapes{\it (Fig. S5)}.  \\

\begin{figure}[!h]
\centering
\includegraphics[width=3.0in,clip]{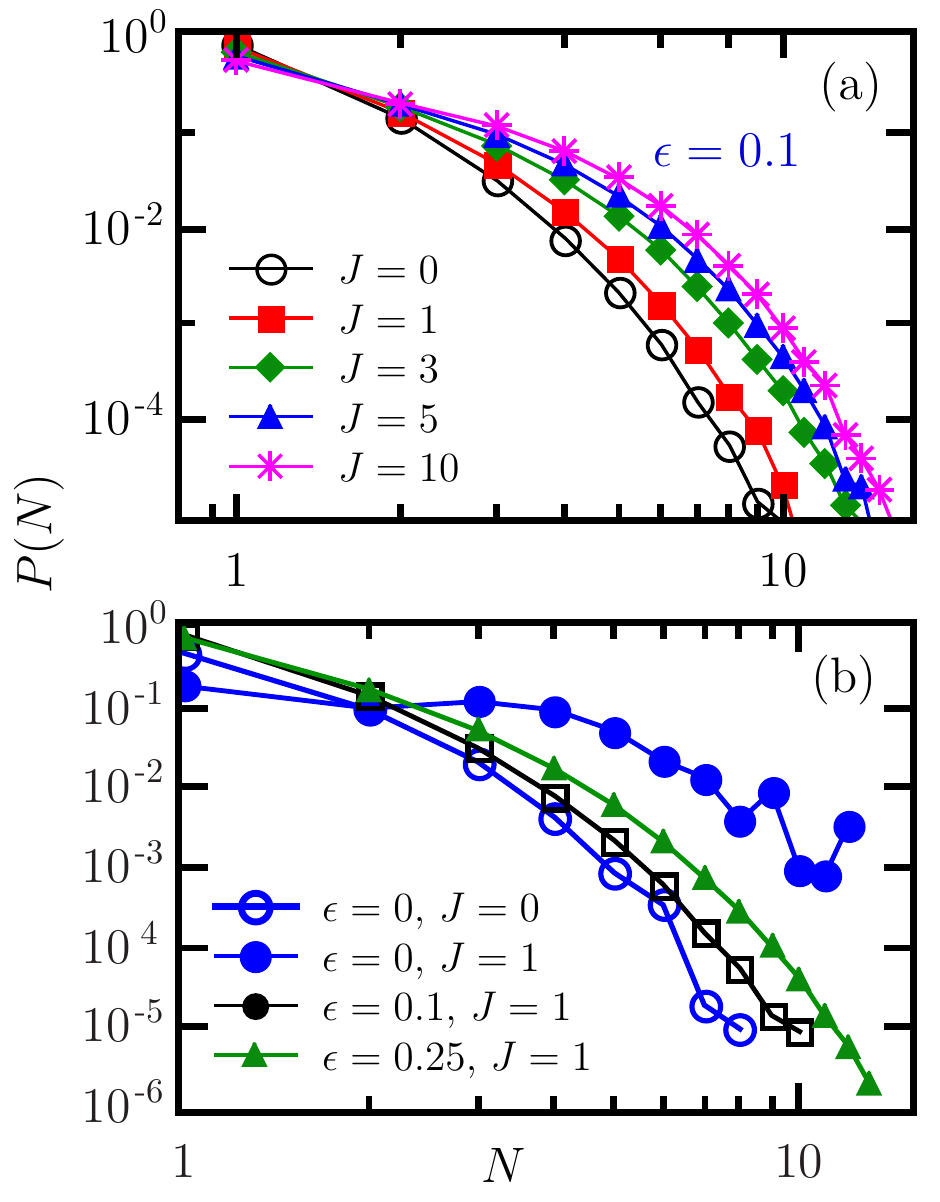}
\caption{\label{fig:clustsize}  {\bf Normalized cluster size distribution of the active species.}  (a)  Cluster size distribution as a function of $J$ at $\epsilon=0.1N/MCS$. (b) Cluster size distribution as a function of activity rate, $\epsilon$. The equilibrium distribution ($\epsilon=0$) for $J=0 \,\&\,1$ has been shown for comparison. Here, $N_{+}^{0}=0.1N$ and $C_{0}=0.8$.}
\end{figure}

\noindent \textit{Effect of activity rate,\,$\epsilon$} :   
The steady state shapes of the active membrane are very sensitive to the rates of activity  and go from tubular to flattened sacs to stomatocyte as the activity is increased (Fig. \ref{fig:NEq-Rate0.1-J0-conf}b).\\
\noindent \textit{Effect of cooperativity, $J$: } The existence of co-operativity between the active species, $J > 0$, promotes the formation of clusters as shown in
 Fig. \ref{fig:clustsize}a, which in turn {\it enhances} the effects of activity and activity-curvature
coupling consistent with earlier studies of membrane mediated aggregation of curvature inducers \cite{Ramaswamy:2000td}. This coupling leads to  the sequence of shapes depicted in Fig.  \ref{fig:NEq-Rate0.1-J0-conf}c even for a small activity rate $\epsilon=0.1$. However, enhancing  activity prevents the formation of larger clusters Fig.  \ref{fig:clustsize}b, a result consistent with Ref. \cite{Turner:2005dt}.
Thus, while for an equilibrium membrane, the critical transition to having large clusters occurs at $J\sim1$ (Fig. \ref{fig:clustsize}a), no large scale clustering occurs at the steady states of an active membrane (with $\epsilon=0.1$) - indeed, though the average domain size increases with $J$, $70\%$ of the active species are still monomeric (Fig. \ref{fig:clustsize}a). 

Cluster size distribution exhibits a dominantly exponential behaviour which is consistent with the low tension regime  reported in \cite{Turner:2005dt} {\it (Fig.  S13 )}.

A detailed steady state phase diagram of the morphology of an active membrane in the $\epsilon-C_{0}$ and $\epsilon - J$ planes are displayed in the supplementary information {\it(Figs. S6 \& S7)}. This shows  that the same ramified or flattened shape can be achieved either by increasing the activity rate or by increasing the activity-curvature coupling of the protein complexes. These phase transitions are characterized by order parameters describing geometrical quantities and composition.

\begin{figure}[!h]
\centering
\includegraphics[width=3.0in]{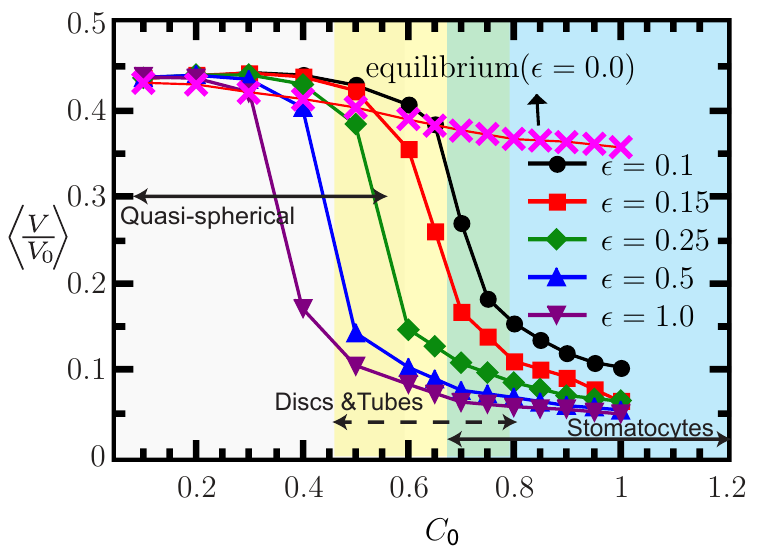}
\caption{\label{fig:volarea-funcrate} Volume($V$) enclosed by a membrane of spherical topology as a function of the spontaneous curvature, $C_0$ with $\kappa=20, J=0.0$ and active field composition $N_{+}^{0}=0.1$N. 
The transition from quasi-spherical, tube, disc and stomatocyte are shown by the various shaded regions for the case of $\epsilon=0.1$.The transition from a quasi-spherical vesicle to other  shapes involves a jump in the enclosed volume and gets sharper as $\epsilon$ increases.  The $\epsilon_{\pm}=0$ curve is with $10\%$ of the vertices having a fixed spontaneous curvature $C_0$.}
\end{figure}

The geometrical order parameter  is the volume or the ratio of the surface area-to-volume, since the surface area of the active vesicle remains roughly constant. The volume, 
scaled by the reference volume $V_{0}\equiv \left(4 \pi/3\right) \left(A/4\pi\right)^{3/2}$,
changes abruptly as the shape transforms from a quasi-spherical conformation to a tubule or  disc, and smoothly goes to zero as the membrane transforms to a stomatocyte (Fig.  \ref{fig:volarea-funcrate}).  This transition can also be observed for other geometrical measures constructed from the eigenvalues of the gyration tensor, like the asphericity and anisotropy {\it (see Fig.  S9 \& S10)}. 
 
This  {\it collapse transition} is of purely non-equilibrium origin, its onset is advanced when $\epsilon$ increases and is absent for an equilibrium vesicle when $\epsilon=0$ (Fig.  \ref{fig:volarea-funcrate}). 

\begin{figure}[!h]
\centering 
\includegraphics[width=3.0in]{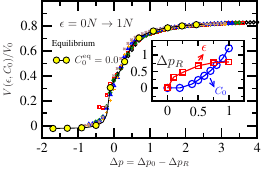}
\caption{\label{fig:collapsed-data} \textit{(Main Panel)}  Data collapse of the non-equilibrium pressure volume curves  ( $\Delta p_{0}$-$V$ curves shown in  Fig. S6)  obtained by shifting the bare osmotic pressure $\Delta p_{0}$ by the DCR dependent $\Delta p_{R}$. Data corresponds to $\kappa=20k_{B}T$, $C_{0}=0.8$, $N_{+}^{0}=0.1N$ with  activity rate in the range $\epsilon=0.0 \rightarrow 10N$. Some points on the equilibrium curve is removed to avoid clutter.  The filled circle represent the data obtained for equilibrium vesicles with $C_0^{\rm eq}=0.0$. \textit{(Inset)} Computed values of $\Delta p_{R}$ as a function of $\epsilon_{\pm}$ and $C_0$. }
\end{figure}

So far we have kept the osmotic pressure $\Delta p$ fixed, we now explore the shape changes on varying $\Delta p$. 
The $\Delta p$-$V$ curves {\it(Fig. S8)}, show a  data collapse when plotted against an activity renormalized osmotic pressure, $\Delta p-\Delta p^{R}(\epsilon)$ (Fig.  \ref{fig:collapsed-data}). This indicates that activity generates a negative pressure that causes the shape changes and eventual collapse to the stomatocyte.

Significantly, the scaling curve (Fig.  \ref{fig:collapsed-data}) coincides with the pressure-volume curve of an equilibrium vesicle with $C_0=0$; further
the renormalized pressure $\Delta p^{R}$ saturates beyond activity rate $\epsilon=1{\rm N/MCS}$ (Fig.  \ref{fig:collapsed-data}(inset)).
This suggests that activity contributes to a negative tension \cite{Solon:2006gk} which leads to a negative Laplace pressure, resulting in the collapse transition. This would imply that the surface of the active vesicle should be highly folded or wrinkled, consistent with our simulations. Likewise, activity gives rise to a modulation of the local spontaneous curvature, which results in a tension and hence a  
renormalized osmotic pressure proportional to $\kappa C_0^2$, as verified  in Fig. \ref{fig:collapsed-data}(inset).
\\

\begin{figure}[!h]
\centering
\includegraphics[width=3.0in,clip]{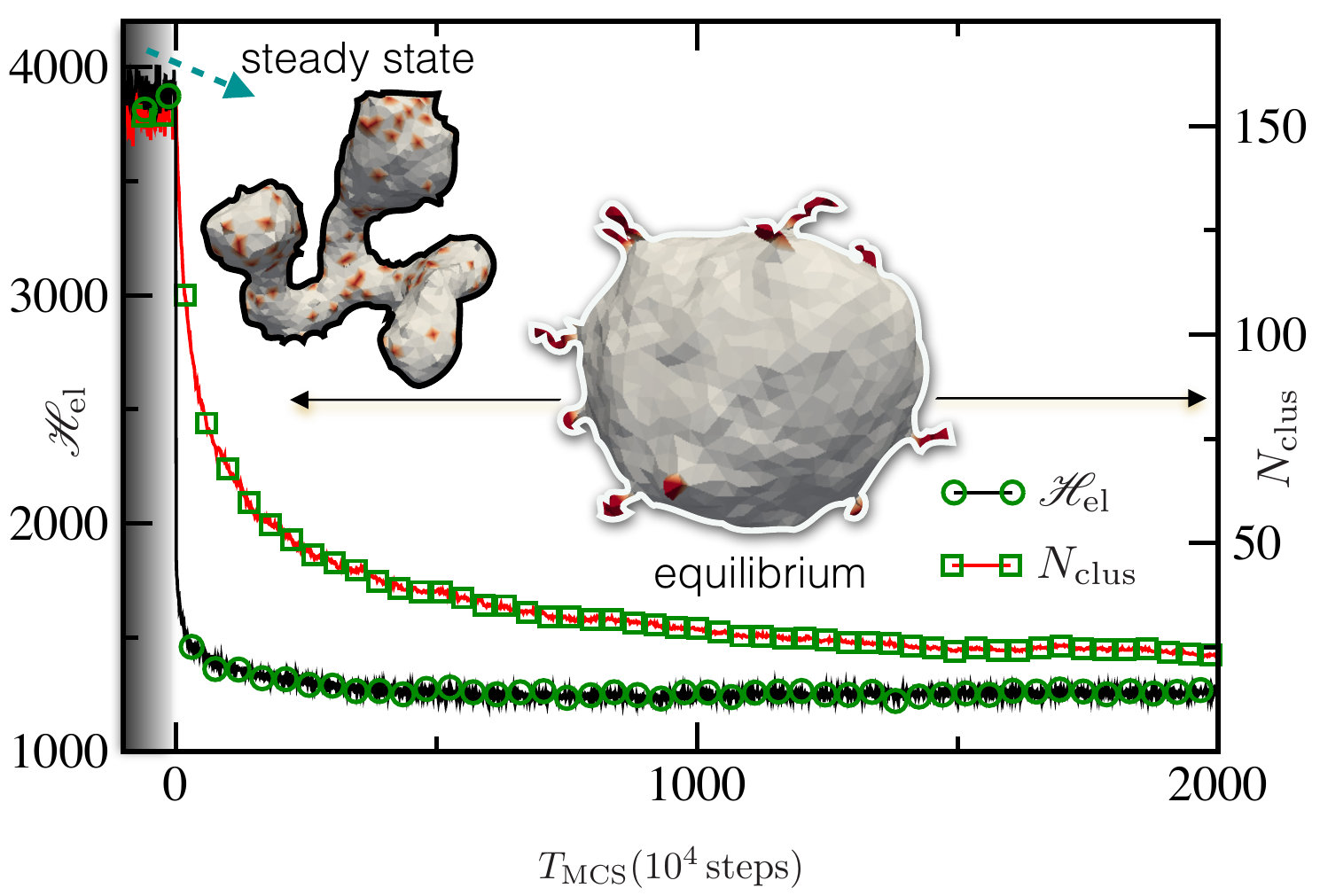}
\caption{\label{fig:steady-state}Activity stabilizes tubular structures for $T_{MCS}<2000$, with $C_0=0.8$, $\kappa=20$, $N_{+}^{0}=0.1$N, $J=1$  and $\epsilon=0.1{\rm N/ MCS}$. On inhibition of activity at $T_{MCS}=2000 \times 10^{4}$ the membrane form buds, characteristic of two phase membranes \cite{SunilKumar:2001cr}, thereby minimizing the elastic energy, $\mathscr{H}_{el}$(filled circles). Further, the regions with the DCR species coarsen  resulting in small number of clusters, $N_{\rm clus}$(open circles), for $T_{MCS}>2000 \times 10^{4}$. }
\end{figure}

\section{Concluding Remarks}
We have shown that the active vesicle subject to non-equilibrium local curvature remodeling processes exhibits morphologies in striking similarity with those displayed by internal cellular organelles. This would suggest that the complex ramified shapes of cell organelles could be a result of non-equilibrium material transport into and out of its bounding membrane. This places {\it nonequilbrium} phenomena at the heart of organelle morphogenesis and appears to have some level of support \cite{Rafelski:2008iy,Chan:2010tt,Shibata:2010fz}. 
This view is to be  contrasted with the view that tubular and flat organelle morphologies are a result of purely {\it equilibrium} forces generated on the membrane by specialized curvature modifying proteins \cite{Ramakrishnan:2012dk,Simunovic:2013ho}.

How does one experimentally  test these seemingly contrasting views ? Fig.  \ref{fig:steady-state} demonstrates how the shape and composition of the active membrane relaxes to equilibrium when the activity $\epsilon$ is abruptly shut off.  Starting from an initial tubular morphology of the active membrane, the shape changes rapidly to give rise to an inflated near-spherical equilibrium morphology. The elastic energy $\mathscr{H}_{\rm el}$ drops exponentially fast  and that the curvature sensing/generating proteins cluster and coarsen, as indicated by a rapid decrease of $N_{\rm clus}$, to form tubular buds (as in  \cite{SunilKumar:2001cr}).  The dynamics of shape change is captured  in the Supplementary Movies M1($J=0$) and M2 ($J=1$). Experimental approaches, in which one monitors the dynamics of shape changes of Golgi compartments using high resolution live-cell imaging
when the agencies of active fission and fusion are suddenly switched off, could help resolve these issues. It is possible however that {\it both} these mechanisms work together - that the specific curvature generating proteins have evolved as embellishments on this organizing non-equilibrium framework, leading to robustness, and efficiency.
We hope the results obtained here will drive further experimental effort to arrive at a deeper understanding of the fundamental issues governing organelle morphogenesis.
 
\begin{acknowledgments}
 This work was supported in part by a grant from the Danish Research Councils (11-107269). MR acknowledges a grant from Simons Foundation. Computations were performed using the HPCE at IIT Madras.
\end{acknowledgments}


\begin{thebibliography}{10}

\bibitem{Rafelski:2008iy}
S.~M. Rafelski and W.~F. Marshall, Nature {\bf 9},  593  (2008).

\bibitem{Marshall:2011kw}
W.~F. Marshall, BMC Biol {\bf 9},  57  (2011).

\bibitem{MartinezMenarguez:2013ek}
J.~A. Mart{\'\i}nez-Men{\'a}rguez, ISRN Cell Biology {\bf 2013},  1  (2013).

\bibitem{Shibata:2009je}
Y. Shibata, J. Hu, M.~M. Kozlov, and T.~A. Rapoport, Annu. Rev. Cell Dev. Biol.
  {\bf 25},  329  (2009).

\bibitem{Rao:2001jl}
M. Rao and S. R~C, Phys. Rev. Lett. {\bf 87},  128101  (2001).

\bibitem{Girard:2004jl}
P. Girard, F. J{\"u}licher, and J. Prost, Eur. Phys. J. E {\bf 14},  387
  (2004).

\bibitem{Klann:2012ew}
M. Klann, H. Koeppl, and M. Reuss, PLoS ONE {\bf 7},  e29645  (2012).

\bibitem{Ispolatov:2013bc}
I. Ispolatov and A. M{\"u}sch, PLoS Comput Biol {\bf 9},  e1003125  (2013).

\bibitem{Alberts:1994}
B. Alberts {\it et~al.}, {\em Molecular Biology of the cell}, 3rd  ed. (Garland
  Publishing, Singapore, 1994).

\bibitem{Wieland:1987tw}
F.~T. Wieland, M.~L. Gleason, T.~A. Serafini, and J.~E. Rothman, Cell {\bf 50},
   289  (1987).

\bibitem{Frost:2009gu}
A. Frost, V.~M. Unger, and P. De~Camilli, Cell {\bf 137},  191  (2009).

\bibitem{DSouzaSchorey:2006gc}
C. D'Souza-Schorey and P. Chavrier, Nature {\bf 7},  347  (2006).

\bibitem{Marks:2001em}
B. Marks {\it et~al.}, Nature {\bf 410},  231  (2001).

\bibitem{Baschieri:2012ia}
F. Baschieri and H. Farhan, Small GTPases {\bf 3},  80  (2012).

\bibitem{Harris:2011db}
K.~P. Harris and J.~T. Littleton, Curr. Biol. {\bf 21},  R841  (2011).

\bibitem{Chavrier:1999ir}
P. Chavrier and B. Goud, Current Opinion in Cell Biology {\bf 11},  466
  (1999).

\bibitem{Turner:2005dt}
M. Turner, P. Sens, and N. Socci, Phys. Rev. Lett. {\bf 95},  168301  (2005).

\bibitem{Staykova:2011cb}
M. Staykova, D.~P. Holmes, C. Read, and H.~A. Stone, Proc. Natl. Acad. Sci.
  U.S.A. {\bf 108},  9084  (2011).

\bibitem{doCarmo:1976}
M.~P. do~Carmo, {\em Differential geometry of curves and surfaces} (Prentice
  Hall, Engelwood Cliffs, New Jersey, 1976).

\bibitem{Gompper:1994cs}
G. Gompper and D. Kroll, Phys. Rev. Lett. {\bf 73},  2139  (1994).

\bibitem{SunilKumar:2001cr}
P. Sunil~Kumar, G. Gompper, and R. Lipowsky, Phys. Rev. Lett. {\bf 86},  3911
  (2001).

\bibitem{Noguchi:2005br}
H. Noguchi and G. Gompper, Proc. Natl. Acad. Sci. U.S.A. {\bf 102},  14159
  (2005).

\bibitem{Ramakrishnan:2010hk}
N. Ramakrishnan, P.~B. Sunil~Kumar, and J.~H. Ipsen, Phys. Rev. E {\bf 81},
  041922  (2010).

\bibitem{Note1}
In the case of membrane bound pumps which switch from their active to inactive
  forms, we take $H_{0i}=C_0\phi _i^2(1+\phi _i)/2$ {\protect \it ( see Sec. S1
  in S.I.)}.

\bibitem{Brandizzi:2013ik}
F. Brandizzi and C. Barlowe, Nature {\bf 14},  382  (2013).

\bibitem{Spang:2013ev}
A. Spang, Cold Spring Harbor Perspectives in Biology {\bf 5},  a013391  (2013).

\bibitem{Im:20030p788}
W. Im, S. Seefeld, and B. Roux, Biophysical Journal {\bf 79},  788  (2000).

\bibitem{Ramaswamy:2000td}
S. Ramaswamy, J. Toner, and J. Prost, Phys. Rev. Lett. {\bf 84},  3494  (2000).

\bibitem{Solon:2006gk}
J. Solon {\it et~al.}, Phys. Rev. Lett. {\bf 97},  098103  (2006).

\bibitem{Chan:2010tt}
Y.-H.~M. Chan and W.~F. Marshall, Organogenesis {\bf 6},  88  (2010).

\bibitem{Shibata:2010fz}
Y. Shibata {\it et~al.}, Cell {\bf 143},  774  (2010).

\bibitem{Ramakrishnan:2012dk}
N. Ramakrishnan, J.~H. Ipsen, and P.~B.~S. Kumar, Soft Matter {\bf 8},  3058
  (2012).

\bibitem{Simunovic:2013ho}
M. Simunovic {\it et~al.}, Biophys. J. {\bf 105},  711  (2013).

\end{thebibliography}


\end{document}